\documentclass[prb,twocolumn,showpacs]{revtex4}
\usepackage{graphicx}
\usepackage{amssymb}
\usepackage{dcolumn}
\usepackage{float}
\usepackage{bm}
\usepackage{multirow}
\usepackage{booktabs}
\usepackage{amsmath}
\usepackage[T1]{fontenc}
\usepackage[utf8]{inputenc}
\usepackage{lmodern}
\usepackage{booktabs}
\usepackage{color}
\usepackage{makecell}
\usepackage{color}

\newcommand{\nn}{\nonumber}

\newcolumntype{M}[1]{>{\centering\arraybackslash}m{#1}}

\newcommand{\e}{\mathrm{e}}

\DeclareMathAlphabet{\bi}{OML}{cmm}{b}{it}

\def\be{\begin{equation}}
\def\ee{\end{equation}}
\def\bearr{\begin{eqnarray}}
\def\eearr{\end{eqnarray}}

\def\ra{\rangle}

\begin{document}
\title{Electrical and optical conductivities of hole gas in $p$-doped bulk III-V semiconductors}
\bigskip
\author{Alestin Mawrie, Pushpajit Halder, Barun Ghosh, Tarun Kanti Ghosh\\
\normalsize
Department of Physics, Indian Institute of Technology-Kanpur,
Kanpur-208 016, India}
\date{\today}
 
\begin{abstract}
We study electrical and optical conductivities of hole gas in $p$-doped
bulk III-V semiconductors described by the Luttinger Hamiltonian.
We provide exact analytical expressions of the Drude conductivity,
inverse relaxation time for various impurity potentials, Drude weight 
and optical conductivity in terms
of the Luttinger parameters $\gamma_1 $ and $\gamma_2$.
The back scattering is completely suppressed as a result of the helicity
conservation of the heavy and light hole states. 
The energy dependence of the relaxation time for the hole states is 
different from the Brooks-Herring formula for electron gas in $n$-doped 
semiconductors.
We find that the inverse relaxation time of heavy holes is much less 
than that of the light holes for Coulomb-type and Gaussian-type impurity
potentials and vice-versa for short-range impurity potential. 
The Drude conductivity increases non-linearly 
with the increase of the hole density. The exponent of the density dependence
of the conductivity is obtained in the Thomas-Fermi limit.
The Drude weight varies linearly with the density even in presence of the
spin-orbit coupling.   
The finite-frequency optical conductivity goes as $\sqrt{\omega}$ and 
its amplitude strongly depends on the Luttinger parameters. 
The Luttinger parameters can be extracted from the optical conductivity
measurement.

\end{abstract}



\maketitle
\section{Introduction}
There is a renewed research interest on various properties of
$p$-doped zinc-blende semiconductors described by the Luttinger 
Hamiltonian \cite{Lutin} for the spin-3/2 valence band,
after the theoretical proposal of intrinsic SHE put forward 
by Murakami et al. \cite{murakami1}. The intrinsic SHE is solely due to 
the presence of the spin-orbit coupling in the bands even in absence 
of any impurities. 
Later, this exotic phenomena has been observed 
experimentally in bulk n-doped semiconductors such as GaAs and InGaAs \cite{she-exp}
as well as in two-dimensional hole gas \cite{she-exp1}. 

A large number of theoretical studies have been carried out on 
$p$-doped bulk III-V semiconductors in recent years.
Quantum mechanical calculations of spin Hall conductivity by defining 
the exact conserved spin current of $p$-doped bulk semiconductors describing by the 
Luttinger Hamiltonian is provided in Ref. \cite{murakami2}.
The wave packet dynamics in the bulk $p$-doped hole gas has been studied
in details \cite{zb1,zb2,zb3}.
A theoretical study of interacting hole gas in
$p$-doped bulk III-V semiconductors has been done using the self-consistent 
Hartree-Fock method\cite{jSc1}.
The dielectric function and beating pattern of the Friedel oscillations of
the bulk hole liquid within the random phase approximation are also studied
\cite{jSc2,jSc3}.
There have been extensive theoretical \cite{fms-t,fms-t1} and 
experimental \cite{fms-e,fms-e1,fms-e2,fms-e3,fms-e4} 
studies on $p$-doped III-V ferromagnetic semiconductors \cite{fms-rev} 
such as GaMnAs and InMnAs with  
the Curie temperature T<170 K.
Recently, the magnetotransport coefficients of the Luttinger Hamiltonian have been studied 
numerically \cite{sdh}.

Studies of the electrical conductivity helps to determine the nature of the
interaction between charge carriers and impurities.
On the other hand, the optical transitions between spin-split Fermi surfaces      
provide the information about the curvature of the complex energy bands.
To the best of our knowledge, 
DC and AC electric fields response to the 3D hole gas in $p$-doped bulk 
semiconductors are lacking. 
The nature of the complex valence bands and values of the Luttinger 
parameters can be extracted from the electrical and optical conductivities.
Therefore, theoretical studies of the electrical and optical conductivities 
would help to analyze the experimental observations. 
Theoretical studies of the Drude and optical conductivities 
of two-dimensional hole gas with $k$-cubic Rashba and Dresselhaus spin-orbit
interactions are given in Refs. \cite{opt4,opt5}.

In this work, we consider hole gas in $p$-doped bulk III-V semiconductors 
described by the Luttinger Hamiltonian subjected to static and time-varying
electric fields and study Drude conductivity, inverse relaxation time, Drude
weight and optical conductivity. 
The exact analytical expressions of the Drude conductivity, inverse 
relaxation times for various impurity potentials, the Drude weight and 
optical conductivity are provided.
We show that the back scattering is completely suppressed due to 
the helicity conservation of the heavy and light hole states. 
The zero-frequency Drude weight is linear 
with respect to the carrier density even in the presence of the spin-orbit 
coupling, due to $k^2$-dependence of the spin-orbit interaction. 
A minimum photon energy is required to trigger the optical transition and
then the optical conductivity grows as $\sqrt{\omega}$ 
($\hbar \omega $ is the photon energy) and cease to zero beyond some critical photon
energy depending on the Luttinger parameters and hole density. 
We show that the Luttinger parameters can be extracted from the optical 
conductivity measurement.

This paper is organized as follows. The basic information of 
the physical system  is described in section II. 
In section III, we present detail calculations of the Drude conductivity and 
the inverse relaxation time. In section IV,
we present results of the Drude weight and the optical conductivity.
An alternate derivation of the optical conductivity is provided in
Appendix.
The summary of this paper is presented in section V.

\section{Basic informations}
 The valence bands of common semiconductors having diamond and 
zinc-blende crystal structures can be accurately
described by the following $6 \times 6$ Luttinger Hamiltonian \cite{Lutin6x6}:
\begin{widetext}
\begin{eqnarray}\label{6by6}
H=\begin{bmatrix}
-P-Q & L & -M & 0 & \frac{1}{\sqrt{2}}L & -\sqrt{2} M\\
L^\dagger & -P+Q & 0 & -M & \sqrt{2}Q & -\sqrt{\frac{3}{2}} L\\
-M^\dagger & 0 & -P+Q & -L & -\sqrt{\frac{3}{2}}L^\dagger & -\sqrt{2} Q\\
0 & -M^\dagger & -L^\dagger & -P-Q & \sqrt{2}M^\dagger & \frac{1}{\sqrt{2}} L^\dagger\\
\frac{1}{\sqrt{2}}L^\dagger & \sqrt{2}Q^\dagger & 
-\sqrt{\frac{3}{2}}L & \sqrt{2}M & - P - \Delta_{\rm so}  & 0\\
-\sqrt{2}M^\dagger & -\sqrt{\frac{3}{2}}L^\dagger & 
-\sqrt{2}Q^\dagger& \frac{1}{\sqrt{2}}L & 0 & - P - \Delta_{\rm so}
\end{bmatrix},
\end{eqnarray}
\end{widetext}
where 
$P=\frac{\gamma_1 \hbar^2}{2m_0}(k_x^2+k_y^2+k_z^2),
Q=\frac{\gamma_2 \hbar^2}{2m_0} (k_x^2+k_y^2-2k_z^2),
L=\frac{\sqrt{3}\gamma_3 \hbar^2}{m_0} (k_x-i k_y)k_z $
and 
$ M = - \frac{\sqrt{3} \hbar^2}{2m_0} 
[\gamma_2(k_x^2-k_y^2)-2i \gamma_3k_xk_y] $.
Here $m_0$ is the bare electron mass and $\Delta_{\rm so} $ being 
the split-off energy. Also, $\gamma_1$, $\gamma_2$ and $\gamma_3$ are 
the dimensionless Luttinger parameters characterizing the valence band 
of the specific semiconductors. The parameters $\gamma_2$ and $\gamma_3$ 
contain the information about the spin-orbit coupling.
The Luttinger parameters $\gamma_1$, $\gamma_2$  and $\gamma_3$ along with 
other parameters of the band structure of various semiconductors are readily 
available in Ref. \cite{parameters}.
The Luttinger parameters in zinc-blende semiconductors are of the order of
the same magnitude.
It implies that the spin-orbit coupling is strong in bulk hole systems,
in comparison to the bulk electron systems.

The split-off energy $\Delta_{\rm so} $ for these semiconductors is of 
the order of few hundred meV. On the other hand, the Fermi energy is of the 
order of few meV for typical hole density ($n_h \sim 10^{23} $ m$^{-3})$. 
Thus, one can safely ignore the split-off band when the Fermi energy is sufficiently 
smaller than the split-off energy and hence the upper-left 
$4 \times 4$ matrix block in Eq. (\ref{6by6}) describes the two upper most 
valence bands (known as heavy hole and light hole bands) approximately.
Within the spherical approximation ($\gamma_3 = \gamma_2$),
the 4 $\times $ 4  Luttinger's Hamiltonian \cite{Lutin} can be written in a 
compact form as
\begin{eqnarray}\label{hamil}
H =\frac{1}{2m_0}\Bigg[\Bigg(\gamma_1+\frac{5}{2}\gamma_2\Bigg){\bf p}^2 - 
2\gamma_2\big({\bf p}\cdot{\bf S}\big)^2\Bigg].
\end{eqnarray}
Here $ {\bf p} $ is the hole momentum operator,
$ {\bf S} $ are the spin-3/2 operators arises from the addition of
$l=1$ orbital angular momentum and $s=1/2$ spin angular momentum.

The spin-3/2 operators are given as
\begin{eqnarray}
S_x=\begin{pmatrix}
0 &\frac{\sqrt{3}}{2} & 0 & 0\\
\frac{\sqrt{3}}{2} & 0 & 1 & 0\\
0 & 1 & 0 & \frac{\sqrt{3}}{2}\\
0 & 0 & \frac{\sqrt{3}}{2}& 0
\end{pmatrix},
\end{eqnarray}

\begin{eqnarray}
S_y=i\begin{pmatrix}
0 &-\frac{\sqrt{3}}{2} &0 &0\\
\frac{\sqrt{3}}{2} &0&-1&0\\
0 &1 &0 & -\frac{\sqrt{3}}{2}\\
0 & 0 & \frac{\sqrt{3}}{2}&0
\end{pmatrix},
\end{eqnarray}

\begin{eqnarray}
S_z=\begin{pmatrix}
\frac{3}{2} & 0 & 0 & 0\\
0 & \frac{1}{2}& 0 & 0\\
0 & 0 &-\frac{1}{2} & 0\\
0 & 0 & 0 &-\frac{3}{2}
\end{pmatrix}.
\end{eqnarray}

The rotationally invariant Hamiltonian $H$
commutes with the helicity operator $ \hat \lambda={\bf k}\cdot{\bf S}/k$
so that its eigenvalues $\lambda= \pm 3/2, \pm 1/2$ are good quantum numbers.
Here $\lambda=\pm 3/2$ and $\lambda= \pm 1/2$  correspond to the heavy hole
and light hole states, respectively. Therefore, the eigenstates of the helicity operator
can be chosen as the eigenstates of the above Hamiltonian. 
The dispersion relations of the heavy and light hole states are given by
\begin{eqnarray}
E_{h/l}({\bf k}) = \frac{(\hbar k)^2}{2m_{h/l}}.
\end{eqnarray}
Here the heavy and light hole masses are 
$ m_{h/l} = m_0/(\gamma_1 \mp 2\gamma_2) $, respectively.
The two-fold degeneracy of heavy and light hole branches is due to the
consequence of the space inversion and time-reversal symmetries of the Luttinger 
Hamiltonian. 
The corresponding eigenstates are given by
\begin{eqnarray}
\psi_{\lambda, {\bf k}}({\bf r}) = \frac{e^{i {\bf k \cdot r}}}{\sqrt{V}}
| \phi_\lambda({\bf k}) \ra,
\end{eqnarray}
where $V$ is the volume of the system. Using the basis of
eigenstates of $S_z$ and parameterizing ${\bf k}$ in terms of spherical polar
coordinates as ${\bf k}=k(\sin\theta\cos\phi,\sin\theta\sin\phi,\cos\theta)$, 
the eigenspinors $| \phi_\lambda({\bf k}) \ra $ for $\lambda=3/2$ and $\lambda=1/2$ 
can be written as 
\begin{eqnarray}
|\phi_{3/2}({\bf k}) \ra =\begin{pmatrix}\label{W32}
\cos^3\frac{\theta}{2}e^{(-3i/2)\phi}\\
\sqrt{3}\cos^2\frac{\theta}{2}\sin\frac{\theta}{2} e^{(-i/2)\phi}\\
\sqrt{3}\cos\frac{\theta}{2}\sin^2\frac{\theta}{2} e^{(i/2)\phi}\\
\sin^3\frac{\theta}{2}e^{(3i/2)\phi}
\end{pmatrix}
\end{eqnarray}
and
\begin{eqnarray}\label{W12}
| \phi_{1/2}({\bf k}) \ra =\begin{pmatrix}
-\sqrt{3}\cos^2\frac{\theta}{2}\sin\frac{\theta}{2}e^{(-3i/2)\phi}\\
\cos\frac{\theta}{2}\Big(\cos^2\frac{\theta}{2}-2\sin^2\frac{\theta}{2}\Big) e^{(-i/2)\phi}\\
\sin\frac{\theta}{2}\Big(2\cos^2\frac{\theta}{2}-\sin^2\frac{\theta}{2}\Big) e^{(i/2)\phi}\\
\sqrt{3}\cos\frac{\theta}{2}\sin^2\frac{\theta}{2}e^{(3i/2)\phi}
\end{pmatrix}.
\end{eqnarray}
The remaining spinors for $\lambda = -3/2$ and $\lambda = -1/2$ 
can easily be obtained from  Eq. (\ref{W32}) and Eq. (\ref{W12})
under the spatial inversion operations 
$\theta\rightarrow\pi-\theta$ and $\phi\rightarrow\pi+\phi$.

In order to study transport properties and electrical/optical conductivities, 
we need to know the ground state properties such as 
the Fermi energy and the corresponding Fermi wave vectors of the two branches 
for a given hole density $n_h$.
Following the standard procedure, the Fermi energy and the corresponding 
Fermi wave vector are respectively given by
\begin{eqnarray}\label{fermi}
E_{\rm f} = \frac{(\hbar {k_{\rm f}^0})^2}{2m_0}\bigg[\frac{\gamma_1^2 -
4\gamma_2^2}{\big[(\gamma_1-2\gamma_2)^{3/2}+(\gamma_1+2\gamma_2)^{3/2}\big]^{2/3}}\bigg]
\end{eqnarray} 
and
\begin{eqnarray}\label{FerVec}
k_{\rm f}^{h/l}=k_{\rm f}^0 
\frac{\sqrt{ \gamma_1\pm2\gamma_2} }{\big[(\gamma_1-2\gamma_2)^{3/2} +
(\gamma_1+2\gamma_2)^{3/2}\big]^{1/3}},
\end{eqnarray}
where $k_{\rm f}^0=(3\pi^2n_h)^{1/3}$. One can easily check that 
$ E_{\rm f} \approx 2 E_{\rm f}^0 = (\hbar k_f^0)^2/m_0$ for typical values 
of the Luttinger parameters. 

The $x$, $y$ and $z$ components of the velocity operator (which will be required to 
calculate the Drude and optical conductivities) are as follows

\begin{widetext}
\begin{eqnarray}\label{vx}
\hat{v}_x({\bf k}) =\begin{pmatrix}
(\mathbb{I}\gamma_1+\gamma_2\sigma_z)k_x-\sqrt{3}\gamma_2\sigma_x k_z &  - 
\sqrt{3}\mathbb{I}\gamma_2 k_-\\
-\sqrt{3}\mathbb{I}\gamma_2 k_+ & (\mathbb{I}\gamma_1 -\gamma_2\sigma_z)k_x
+ \sqrt{3}\gamma_2\sigma_x k_z
\end{pmatrix},
\end{eqnarray}

\begin{eqnarray}
\hat{v}_y({\bf k}) =\begin{pmatrix}
(\mathbb{I}\gamma_1 +\gamma_2\sigma_z)k_y-\sqrt{3}\gamma_2\sigma_y k_z & 
i\sqrt{3}\mathbb{I}\gamma_2 k_-\\
-i\sqrt{3}\mathbb{I}\gamma_2 k_+ & (\mathbb{I}\gamma_1 -\gamma_2\sigma_z)k_y + 
\sqrt{3}\gamma_2\sigma_y k_z
\end{pmatrix}
\end{eqnarray}
and 
\begin{eqnarray}
\hat{v}_z({\bf k}) = \begin{pmatrix}
(\mathbb{I}\gamma_1 -\gamma_2\sigma_z)k_z-\sqrt{3}\gamma_2(k_x\sigma_x+k_y\sigma_y) & \mathbb{O}\\
\mathbb{O} & (\mathbb{I}\gamma_1 +\gamma_2\sigma_z)k_z+\sqrt{3}\gamma_2(k_x\sigma_x+k_y\sigma_y) 
\end{pmatrix}.
\end{eqnarray}
\end{widetext}
Here $\mathbb{I} $ is a $2\times2$ identity matrix, $\sigma_{x,y,z}$ 
are the Pauli's $2\times 2$ matrices and $\mathbb{O}$ represents a $2\times2$ null matrix.

\section{Drude Conductivity and Inverse relaxation time}

\subsection{Drude conductivity}
With the application of weak DC electric field along $x$-direction ${\bf E} = E_x \hat x $,
the hole current density is $ J_h = \sigma_{xx} E_x$ with $\sigma_{xx} $ being the Drude 
conductivity. 
Within the semi-classical Boltzmann theory and the relaxation time approximation, 
the general expression of the Drude conductivity at very low
temperature is given by \cite{mermin}
\begin{eqnarray}
\sigma_{xx}=\frac{e^2}{(2\pi)^3}\sum_\lambda \int  d^3k
\langle \hat{v}_x({\bf k})\rangle^2_{_{\lambda}} \, \tau_\lambda({\bf k}) \, 
\delta[E_\lambda({\bf k})-E_{\rm f}],
\end{eqnarray}
where $\hat{v}_x({\bf k})$ is the $x$-component of the velocity operator and
$\tau_\lambda({\bf k})$ is
the relaxation time.

The expectation values of the velocity operator $\hat v_x $ with respect to the 
heavy hole and light hole states are
$ \langle \hat{v}_x({\bf k})\rangle_{ h} = \hbar k_x/m_h $ and
$ \langle \hat{v}_x({\bf k})\rangle_{ l} = \hbar k_x/m_l $.
Using these results, the final expression of the Drude conductivity is 
obtained as
\begin{eqnarray}\label{drude}
\sigma_{xx}=\frac{n_h e^2 }{m_0}\bigg[
\frac{\tau_h \sqrt{\gamma_1+2\gamma_2} + \tau_l \sqrt{\gamma_1-2\gamma_2} }
{(\gamma_1-2\gamma_2)^{3/2} + (\gamma_1+2\gamma_2)^{3/2}}\bigg](\gamma_1^2 -4 \gamma_2^2).
\end{eqnarray}
Here $ \tau_h \equiv \tau_h(E_{\rm f}) $ and  $ \tau_l \equiv \tau_l(E_{\rm f}) $.
Now we need to calculate relaxation time $ \tau_h  $ and $ \tau_l $, which will be shown
in the next section. 
In absence of the atomic spin-orbit coupling (i.e. $\gamma_2 = 0 $), the Drude 
conductivity reduces to the known result. The Drude conductivity varies 
non-linearly with the Luttinger's parameters due to the presence of the
spin-orbit coupling $\gamma_2$.

\subsection{Inverse relaxation time}
In this section, we derive the expression of the inverse relaxation time for
heavy hole and light hole bands. Within the semi-classical Boltzmann theory,
the most general expression of the inverse relaxation time for a given band
is given as\cite{mermin},
\begin{equation}
\frac{1}{\tau_\lambda(k)} = V 
\int \frac{d^3k^{\prime} }{(2\pi)^3} W_{ {\bf k}{\bf k}^{\prime} }^{\lambda } 
(1 - \cos\theta^{\prime}), 
\end{equation}
where $ \theta^{\prime} $ is the angle between the vectors
${\bf k} $ and ${\bf k}^{\prime} $ and 
the intra-band transition rate between the states $|{\bf k},\lambda\rangle$ and  
$|{\bf k^\prime},\lambda \rangle$ is 
$$
W_{ {\bf k}{\bf k}^{\prime}}^{\lambda}=\frac{2\pi}{\hbar}N_{\textrm{\rm imp}} 
\big|\big\langle{\bf k^{\prime},\lambda} 
\big|V_{\textrm{imp}}^{\lambda}({\bf r})\big|{\bf k,\lambda}\rangle\big|^2 
\delta(E_{\lambda}({\bf k})-E_{\lambda}({\bf k^{\prime}})).
$$
Here, $N_\textrm{\rm imp}$ is the number of impurities present in the system
and $ V_{\textrm{imp}}^{\lambda}({\bf r}) $ is the impurity potential. 
The matrix element
$\langle{\bf k^\prime},\lambda \big| V_{\textrm{imp}}^{\lambda} ({\bf r})\big| 
{\bf k},\lambda\rangle$ is given by
$
\langle{\bf k}^{\prime},\lambda\big| V_{\textrm{imp}}^{\lambda} ({\bf r}) \big|
{\bf k},\lambda\rangle =
V_{\rm imp}^{\lambda}({\bf q})\phi_\lambda^\dagger({\bf k}^\prime) \phi_\lambda ({\bf k})/V,
$
where $ q = |{\bf k} - {\bf k}^{\prime}| $ being the change in the wave vector and
$V_{\rm imp}^\lambda({\bf q}) $ is the Fourier transform of 
the impurity potential $V_{\rm imp}^{\lambda}({\bf r}) $.
Therefore, the transition rate is given by 
\begin{eqnarray}
W_{ {\bf k}{\bf k}^{\prime} }^{\lambda }= \frac{2\pi n_{\textrm{imp}} }{\hbar \, V }
\big|V_{\rm imp}^\lambda({\bf q})\big|^2
\big|\phi_\lambda^\dagger({\bf k}^\prime)\phi_\lambda({\bf k})\big|^2
\delta(E_{\lambda}({\bf k})-E_{\lambda}({\bf k^{\prime}}) \nn.
 \end{eqnarray}
Here, $ n_{\rm imp} = N_{\rm imp}/V $ is the impurity density and 
the square of the wave function overlaps are given by 
$$
| \phi_{h}^{\dagger}({\bf k^\prime}) \phi_{h}({\bf k}) |^2 =
\frac{(1 + \cos \theta^{\prime})^3}{8} 
$$ and
$$
 | \phi_{l}^{\dagger}({\bf k^\prime})  \phi_{l}({\bf k}) |^2 =
\frac{(1+ \cos \theta^{\prime}) (3 \cos \theta^{\prime} -1 )^2}{8}. 
$$
It is interesting to note that the square of the wave function overlaps 
exactly vanish at $\theta^\prime = \pi$ and do not contribute to the
scattering rates. It implies that the backscattering is completely suppressed.
This can easily be understood from the helicity conservation.
Since the helicity is a conserved quantity, the charge carriers
can not change its momentum after scattering.
This is similar to the absence of backscattering in monolayer graphene as
a result of the pseudospin conservation \cite{ando,ando1}.
Moreover, the light hole scattering rate ($1/\tau_l$) is also suppressed
at $ \theta^{\prime} = \cos^{-1}(1/3) $.
It should be mention here that the suppression of scattering in certain directions
is independent of choice of the impurity potential.

We shall consider two long-range and one short-range impurity potentials to 
calculate $1/\tau_{\lambda} $ and $\sigma_{xx}$.\\ 
i) {\bf Coulomb impurity potential}: First we consider screened Coulomb-type 
impurity potential, 
$V_{\rm imp}^{\lambda}({\bf r}) = [Ze^2/(4\pi\epsilon)] \e^{-k_s^\lambda r}/r$, 
where $ \epsilon$ is the static dielectric constant  
and $k_{s}^\lambda$ is the Thomas-Fermi 
screening wave-vector, given by
$ {k_{s}^{\lambda}}= \sqrt{4 k_{\rm f}^{\lambda}/(\pi a_B^{\lambda})} $ with 
$ a_B^{\lambda} = 4\pi \epsilon \hbar^2/(Z e^2 m_{\lambda}) $ is the effective
Bohr radius. 
The Fourier transform of the screened Coulomb-type potential is
$V_{\rm imp}^\lambda({\bf q})=\frac{Ze^2}{\epsilon}\frac{1}{(k^\lambda_{s})^2+q^2}$.
Using the above results, the expressions of the inverse relaxation time 
for heavy hole and light hole states are
\begin{widetext}

\begin{equation}
\frac{1}{\tau_h({\bf k}) } =  \frac{n_{\rm imp} m_h}{16 \pi \hbar^3}
\Big[\frac{Ze^2}{\epsilon}\Big]^2  k
\int_0^{\pi} d\theta^\prime \sin \theta^\prime
\frac{(1-\cos\theta^\prime)(1+\cos\theta^\prime)^3}{[(k_{s}^{h})^2 + 2k^2(1-\cos\theta^\prime)]^2}
\end{equation}
and 
\begin{equation}
\frac{1}{\tau_{l}({\bf k}) } =  \frac{n_{\rm imp} m_l}{16 \pi \hbar^3}
\Big[\frac{Ze^2}{\epsilon}\Big]^2  k
\int_0^{\pi} d\theta^\prime \sin\theta^\prime
\frac{(1-\cos^2\theta^\prime)(3\cos\theta^\prime -1)^2}{[(k_{s}^{l})^2 + 
2k^2(1-\cos\theta^\prime)]^2}.
\end{equation}

\end{widetext}
Performing $\theta^\prime$ integrals and expressing in terms of the dimensionless energy
variable $ E $ (in units of $E_{\rm f}^0 $), the inverse relaxation times of heavy 
and light holes are obtained as
\begin{widetext}
\begin{eqnarray}
\frac{\tau_0}{\tau_h(E)} & = &
\frac{(\gamma_1 + 2\gamma_2)\sqrt{\gamma_1^2-4\gamma_2^2} E^{-9/2}}
{(\gamma_1-2\gamma_2)^{3/2}+(\gamma_1+2\gamma_2)^{3/2} }
\Big[ (1+ d_h E)(1+4d_h E)^2 
\log(1+4 d_h E) - \frac{4}{3} d_h E (3 + d_h E (21 + 34 d_h E)) \Big] \nn \\
\frac{\tau_0}{\tau_l(E)} & = &  
\frac{(\gamma_1 - 2\gamma_2)\sqrt{\gamma_1^2 -4 \gamma_2^2} E^{-9/2}}
{(\gamma_1-2\gamma_2)^{3/2}+(\gamma_1+2\gamma_2)^{3/2} }
\Big[ (3+ 4d_l E)(3 + d_l E (11 + 4 d_l E)) 
\log(1+4 d_l E)
- 4 d_l E (3 + 2d_l E) (3 + 7 d_l E) \Big] \nn. 
\end{eqnarray}
\end{widetext}
Here, $1/\tau_0 = \Big[\frac{Ze^2}{4\pi \epsilon} \Big]^5  
\frac{n_{\rm imp} m_0}{\pi^2(\hbar E_{\rm f}^0)^3} $ and
$ d_{\lambda} = \big[k_{\rm f}^0/k_s^{\lambda}\big]^2(\gamma_1 \mp 2\gamma_2)^{-1}$.
The energy variation of $1/\tau (E) $ for hole system is different from 
that of the electrons in $n$-doped bulk semiconductors described by the 
Brooks-Herring formula \cite{brooks}.

\begin{figure}[htbp]
\begin{center}\leavevmode
\includegraphics[width=95mm,height=100mm]{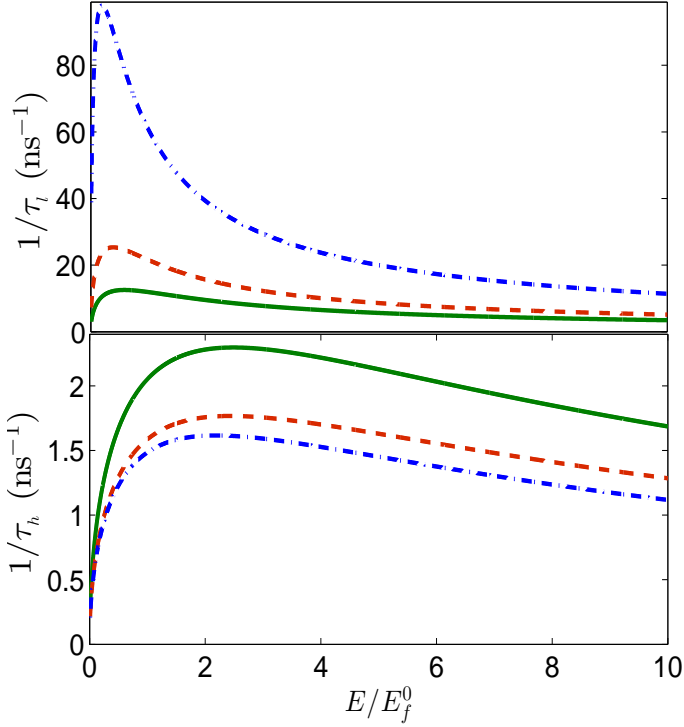}
\caption{Plots of the inverse relaxation time $1/\tau_{h/l}$ versus energy for the 
Coulomb impurity potential for three different semiconductors: 
AlSb (solid green), GaAs (dashed red) and InAs (dashed-dot blue).}
\label{Fig1}
\end{center}
\end{figure}

For various plots, we have taken
$ n_{\rm imp} = 10^{21} $ m${}^{-3}$ and $n_h=5\times10^{23}$ m${}^{-3}$.
In Fig. (1), we show the variations of $1/\tau_{h/l}(E) $ with the energy 
$E$ (in units of $E_{\rm f}^0$) for three different semiconductors. 
It clearly shows that $ \tau_h(E) > \tau_l(E)$ due to huge mass
difference between heavy and light holes. Using the results of the relaxation times
at the Fermi energy, the variation of the Drude conductivity with the hole
density is shown in Fig. 2.

\begin{figure}[htbp]
\begin{center}\leavevmode
\includegraphics[width=95mm,height=70mm]{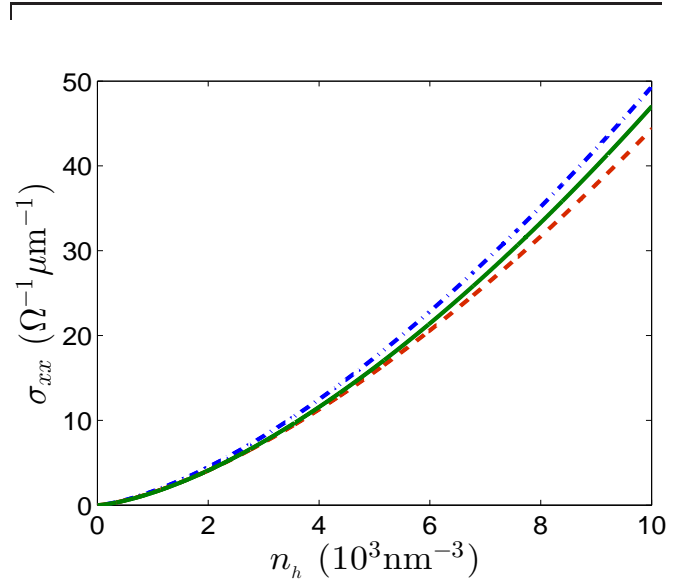}
\caption{Plots of the Drude conductivity $\sigma_{xx} $ versus hole density 
$n_h$ for the Coulomb potential  for three different
semiconductors: AlSb (solid green), GaAs (dashed red) and InAs (dashed-dot blue).}
\label{Fig1}
\end{center}
\end{figure}

In the Thomas-Fermi limit, $k_s^{\lambda} \gg k_{\rm f}^{\lambda} $,
the Coulomb potential in momentum space can be approximated as
$ V_{\rm imp}^{\lambda}({\bf q}) \simeq Ze^2/(\epsilon [k_s^{\lambda}]^2)$.
Within the Thomas-Fermi approximation, the inverse relaxation times are
given by
\begin{equation}
\frac{1}{\tau_{\lambda}(E_{\rm f})} \simeq 
a_{\lambda} \frac{n_{\rm imp} m_{\lambda} }{16 \pi \hbar^3}
\Big[\frac{Ze^2}{\epsilon}\Big]^2  \frac{k_{\rm f}^{\lambda}}{(k_s^{\lambda})^4},
\end{equation}
where $ a_h = 8/5$ and $a_l = 56/15$.
The ratio between $\tau_h $ and $\tau_l$ is
\begin{equation}
\frac{\tau_h}{\tau_l} = \frac{7}{3} \Big(\frac{m_h}{m_l}\Big)^{3/2}
= \frac{7}{3}
\Big[\frac{\gamma_1 +2 \gamma_2}{\gamma_1 -2 \gamma_2}\Big]^{3/2}.
\end{equation}
This ratio depends solely on the Luttinger parameters and clearly indicates that
$\tau_h > \tau_l$.
The Drude conductivity in Thomas-Fermi regime is given by
\begin{equation}
\sigma_{xx} \simeq \frac{n_h e^2 \tau_B}{m_0} \frac{16 k_{\rm f}^0 }{\pi^3 a_B^2 n_{\rm imp} }
\frac{a_{h}^{\prime}(\gamma_1 + 2\gamma_2)^2 + a_{l}^{\prime}(\gamma_1 - 2\gamma_2)^2}
{\big[ (\gamma_1-2\gamma_2)^{3/2} + (\gamma_1 + 2\gamma_2)^{3/2} \big]^{4/3} } \nn,
\end{equation}
where $ \tau_B = (4\pi \epsilon/Ze^2)^2(\hbar^3/m_0)$,
$ a_B = 4\pi \epsilon \hbar^2/(Z e^2 m_0) $ and
$a_{h/l}^{\prime} = 1/a_{h/l} $.
The above equation depicts that the density dependence of the Drude conductivity 
due to the Coulomb impurity potential is $n_h^{4/3}$.

ii) {\bf Gaussian impurity potential}: The Gaussian impurity potential
is taken as
$ V^{\lambda}({\bf r}) = V_0 e^{-(k_s^{\lambda} r)^2/2} $ with $V_0$ is the strength of the
potential. Its Fourier transform is 
$ V^{\lambda}({\bf q}) = V_0 (\sqrt{2\pi}/k_s^{\lambda})^3 
e^{-q^2/(\sqrt{2} k_s^{\lambda})^2} $.
Following the same method as described above, the inverse relaxation times for heavy and light
holes are given by

\begin{equation}
\frac{1}{\tau_h({\bf k}) } =  \frac{n_{\rm imp} \pi^2 m_h V_0^2 k}{2 \hbar^3 (k_{s}^{h})^6}  
\int_0^{\pi} d\theta^\prime \sin^3\theta^\prime (1+\cos\theta^\prime)^2 e^{-(q/k_s^{h})^2} \nn
\end{equation}
and
\begin{equation}
\frac{1}{\tau_{l}({\bf k}) } =  \frac{n_{\rm imp} \pi^2 m_l V_0^2 k}{2\hbar^3 (k_{s}^{l})^6} 
\int_0^{\pi} d\theta^\prime \sin^3\theta^\prime (3\cos\theta^\prime -1)^2  e^{-(q/k_s^{l})^2} \nn,
\end{equation}
where $ q^2 = 2k^2(1-\cos \theta^\prime) $.
After performing the angular integral, the inverse relaxation 
time for heavy and light holes are given by
\begin{eqnarray}
\frac{\tau_0}{\tau_h(E)} & = & 
\frac{ \sqrt{\gamma_1 - 2\gamma_2}(\gamma_1^2-4\gamma_2^2) E^{-9/2} }
{[(\gamma_1-2\gamma_2)^{3/2}+(\gamma_1+2\gamma_2)^{3/2}]^{2/3} } A_h(E) \nn \\
\frac{\tau_0}{\tau_l(E)} & = & 
\frac{ \sqrt{\gamma_1 + 2\gamma_2}(\gamma_1^2 -4 \gamma_2^2) E^{-9/2} }
{[(\gamma_1-2\gamma_2)^{3/2}+(\gamma_1+2\gamma_2)^{3/2}]^{2/3} }  B_h(E) \nn.
\end{eqnarray}
Here, $A_h(E) =  3(1 + d_h E)e^{-4d_h E}  + 
( - 3 +  d_h E (9 + 4 d_h E(-3 + 2d_h E))) $,
$ B_h(E) =- 27 + 45 d_l E - 28 (d_l E)^2  + (2 d_l E)^3 +
[27 + 63 d_l E + (8d_l E)^2 + 32 (d_l E)^3)] e^{-4d_l E}  $ and
$1/\tau_0 = \Big[\frac{Ze^2}{4\pi \epsilon} \Big]^2
\frac{n_{\rm imp} V_0^2}{\sqrt{32 m_0} (E_{\rm f}^0)^{7/2} } $.

The variations of $1/\tau_{h/l}(E) $ versus energy $E/E_{\rm f}^0$ for the Gaussian 
impurity potential for three different semiconductors are shown in Fig. 3.
It clearly shows that $ \tau_h(E) > \tau_l(E)$ due to huge mass
difference between heavy and light holes. Using the results of the relaxation time
at the Fermi energy, the variation of the Drude conductivity with the hole
density for Gaussian scattering potential is shown in Fig. 4.
\begin{figure}[htbp]
\begin{center}\leavevmode
\includegraphics[width=90mm,height=105mm]{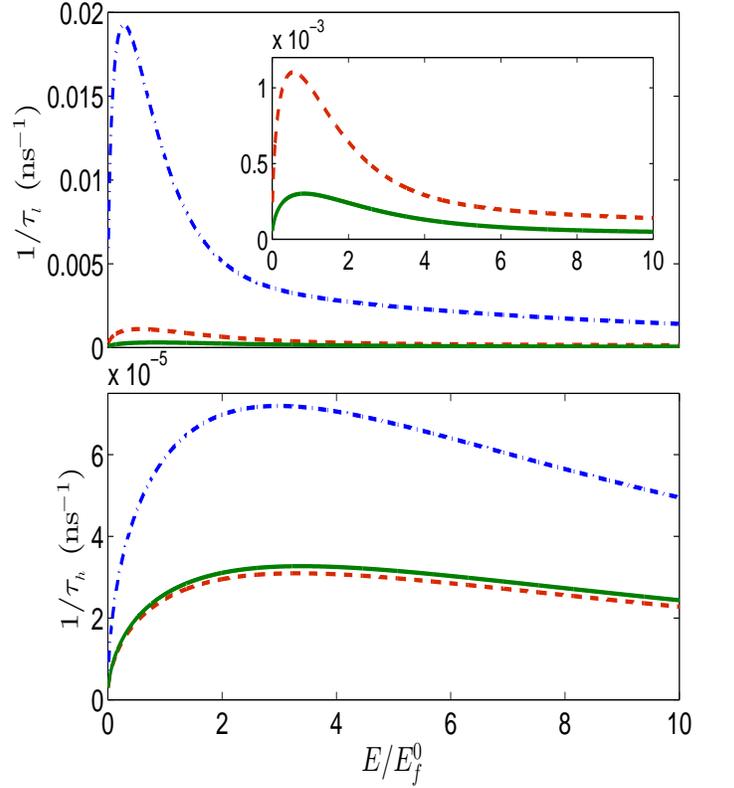}
\caption{Plots of the inverse relaxation time $1/\tau_{h/l}$ versus $E/E_{\rm f}^0$
for the Gaussian impurity potential for three different semiconductors:
AlSb (solid green), GaAs (dashed red) and InAs (dashed-dot blue). Here, we have
taken $ V_0 = 0.1 $ meV.}
\label{Fig1}
\end{center}
\end{figure}
In the Thomas-Fermi limit, $k_s^{\lambda} \gg k_{\rm f}^{\lambda} $,
the Gaussian impurity potential in momentum space can be approximated as
$ V_{\rm imp}^{\lambda}({\bf q}) \simeq V_0 (\sqrt{2\pi}/k_s^{\lambda})^3$.
Within the Thomas-Fermi approximation, the inverse relaxation times at the Fermi
level are given by
\begin{equation}
\frac{1}{\tau_{\lambda}(E_{\rm f})} \simeq
a_{\lambda} \frac{n_{\rm imp} \pi^2 m_{\lambda} V_0^2 
k_{\rm f}^{\lambda} }{2 \hbar^3 (k_{s}^{\lambda})^6}.
\end{equation}
The ratio between $\tau_h $ and $\tau_l$ is
\begin{equation}
\frac{\tau_h}{\tau_l} = \frac{7}{3} \Big(\frac{m_h}{m_l}\Big)^{3}
= \frac{7}{3}
\Big[\frac{\gamma_1 +2 \gamma_2}{\gamma_1 -2 \gamma_2}\Big]^{3}.
\end{equation}
This ratio depends solely on the Luttinger parameters and clearly indicates that
$\tau_h > \tau_l$. Moreover, the ratio $\tau_h/\tau_l$ for the Gaussian impurity
potential is large as compared to the ratio for the Coulomb impurity potential.
The Drude conductivity in the Thomas-Fermi regime is given by
\begin{eqnarray}
\sigma_{xx} & = &
\frac{n_h e^2 \tau_B}{m_0} \frac{256 \hbar^2 E_{\rm f}^0}{\pi^5 V_0^2 n_{\rm imp} m_0 a_B^5}
\nn \\
& \times & 
\frac{ a_{h}^{\prime} (\gamma_1 + 2\gamma_2)^{7/2} + 
a_{l}^{\prime} (\gamma_1 - 2\gamma_2)^{7/2} }
{[(\gamma_1 + 2\gamma_2)^{3/2} + ( \gamma_1 - 2\gamma_2)^{3/2}]^{5/3} 
(\gamma_1^2 - 4 \gamma_2^2)} \nn.
\end{eqnarray}

The density dependence of the Drude conductivity
due to the Gaussian impurity potential is $n_h^{5/3}$.
\begin{figure}[htbp]
\begin{center}\leavevmode
\includegraphics[width=90mm,height=70mm]{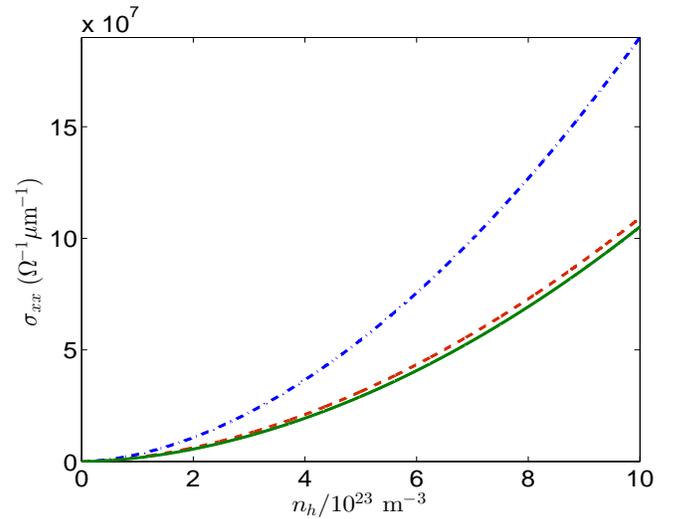}
\caption{Plots of the Drude conductivity $\sigma_{xx} $ versus hole density
$n_h$ for the Gaussian potential for three different
semiconductors: AlSb (solid green), GaAs (dashed red) and InAs (dashed-dot blue).
Here, we have taken $V_0 = 0.1 $ meV.}
\label{Fig1}
\end{center}
\end{figure}

iii) {\bf Short-range impurity potential}: The short-range impurity potential
is considered as $ U({\bf r}) = U_0 \delta({\bf r} - {\bf r}_i) $,
with $U_0$ has the dimension of energy times volume and 
${\bf r}_i$ is the position of the $i$-th impurity. 
Following the same method, the energy variation of the inverse relaxation times 
for short-range impurity potential are obtained as
\begin{eqnarray}
\frac{1}{\tau_h(E)} & = & \frac{m_h^{3/2}n_{\rm imp} U_0^2}{10\pi\hbar^4}\sqrt{2E},\\
\frac{1}{\tau_l(E)} & = & \frac{7m_l^{3/2}n_{\rm imp} U_0^2}{30\pi\hbar^4}\sqrt{2E}.
\end{eqnarray}
The ratio between $\tau_h $ and $\tau_l$ is
$ \tau_h/\tau_l = (7/3) (m_h/m_l)^{3/2}
= (7/3) [(\gamma_1 +2 \gamma_2)/(\gamma_1 -2 \gamma_2)]^{3/2}.$
It shows that $ \tau_h < \tau_l$ and is just opposite to the long-range 
impurity potentials cases. The Drude conductivity is given by
\begin{eqnarray}
\sigma_{xx}= \frac{a_0 e^2\hbar^3
n_h^{2/3} }{ n_{\rm imp}m_0 U_0^2}
\frac{(5 \gamma_1 + 4 \gamma_2) }
{\big[(\gamma_1-2\gamma_2)^{3/2}+(\gamma_1+2\gamma_2)^{3/2}\big]^{2/3}}, \nn
\end{eqnarray}
with $a_0 = 20 \pi/(3 (3\pi^2)^{1/3})$.
The Drude conductivity for the short-range potential varies with the 
carrier density as $n_h^{2/3}$.

Keeping in mind that $ E_{\rm f} \approx  2 E_{\rm  f}^0 $, the
$1/\tau_l $ increases with the energy $E$, peaks at certain value 
$E \ll E_{\rm f}$, and then decreases rapidly while $E$ approaches to $E_{\rm f} $ 
for the Coulomb as well as the Gaussian impurity potentials. 
On the other hand, $1/\tau_h $ increases with energy $E$, peaks in and around 
$E_{\rm f} $ and then decreases slowly when $E \gg E_{\rm f} $ for the
Coulomb and the Gaussian impurity potentials.
For the short-range impurity potential, $1/\tau_{h/l} $ always increases with 
the energy as $\sqrt{E} $.

\section{Drude weight and Optical conductivity}
An oscillating electric field with zero-momentum $ {\bf E} = E_x e^{i \omega t} \hat {\bf x} $ 
is applied on the spin-split hole gas in $p$-doped bulk III-V semiconductors.
The complex charge conductivity is given by
\begin{eqnarray}
\Sigma_{xx}(\omega) = \sigma_D(\omega) + \sigma_{xx}(\omega).
\end{eqnarray}
Here $ \sigma_D(\omega) = 
\sum_{\lambda} \sigma_{xx}^{\lambda}/(1-i \omega \tau_{\lambda}) $ is the dynamic
Drude conductivity arises from the intra-band transitions, with
$\sigma_{xx}^{\lambda} $ being the static Drude conductivity which is
derived in the previous section.
Also, $ \sigma_{xx}(\omega) $ is the complex conductivity arises from the
interband optical transitions between heavy hole and light hole states.
The absorptive parts of the optical transitions correspond to the real parts
of the complex optical conductivities $\sigma_D $ and $\sigma_{xx}(\omega)$.
The minima in the experimentally observed spectra correspond to the peaks
in the real part of the conductivities. \\

{\bf Drude weight}:
The real part of the dynamic Drude conductivity is 
${\rm Re} \, [\sigma_D(\omega)] = D_w \delta(\omega)$, 
where $D_w = \pi \sum_{\lambda} \sigma_{xx}^{\lambda}/\tau_{\lambda} $ is 
called the zero-frequency Drude weight, whose peak is centered around $\omega=0$.
Using Eq. (\ref{drude}) for $\sigma_{xx} $, the Drude weight is given by
\begin{eqnarray}\label{drudeP}
D_{w}=\frac{\pi n_he^2}{m_0}\bigg[\gamma_1+2\gamma_2 
\frac{(\gamma_1-2\gamma_2)^{3/2}-(\gamma_1+2\gamma_2)^{3/2}}{(\gamma_1-2\gamma_2)^{3/2}
+ (\gamma_1+2\gamma_2)^{3/2}}\bigg].
\end{eqnarray}
The Drude weight is linearly varying with the carrier density $n_h$ as
expected since the full Hamiltonian is quadratic in momentum. 

{\bf Optical conductivity}:
Within the framework of linear response theory, the Kubo formula for the 
optical conductivity is given by
\begin{eqnarray}
\sigma_{\mu \nu}(\omega)=\frac{1}{\hbar(\omega-i \eta)}\int_0^\infty dt\textrm{ } 
\e^{(i(\omega+i\eta)t)}\langle[\hat j_\mu(t), \hat j_\nu(0)]\rangle,
\end{eqnarray}
where $(\mu,\nu=x,y,z)$, $\hat j_\mu =e  \hat v_\mu$ is the charge 
current density operator. The quantity 
\begin{eqnarray}
\langle[\hat j_\mu(t), \hat j_\nu(0)]\rangle&=&\sum_{\lambda,\lambda^\prime}
\int d^3{\bf k}f(E_\lambda({\bf k})) 
\langle {\bf k}\lambda^\prime\big|[\hat j_\mu(t), \hat j_\nu(0)]\big|{\bf k}\lambda\rangle\nonumber\\
&=&\sum_{\lambda,\lambda^\prime} 
\Big[f(E_\lambda({\bf k}))-f(E_{\lambda^\prime}({\bf k}))\Big]\nonumber\\
&\times&\e^{i(E_\lambda({\bf k})-E_{\lambda^\prime}({\bf k}))t/\hbar} 
j_\mu^{\lambda,\lambda^\prime}j_\nu^{\lambda^\prime,\lambda}
\end{eqnarray}
with $f(\epsilon)=1/(\exp[\beta(\epsilon-\epsilon_f)] + 1)$ being the Fermi 
distribution function and $j_\mu^{\lambda,\lambda^\prime}=ev_\mu^{\lambda,\lambda^\prime}=
e\langle \lambda| \hat v_\mu| \lambda^\prime\rangle$.
Using the above mention results, the absorptive part of 
$\sigma_{xx}(\omega)$ is simplified to 

\begin{widetext}

\begin{eqnarray}\label{OpT}
\textrm{Re} \,[ \sigma_{xx}(\omega) ] & = &-\frac{e^2}{8\pi^2\omega}
\sum_{\lambda,\lambda^\prime}\int d^3  k
\Big[f(E_{\lambda}({\bf k}))-f(E_{\lambda^\prime}({\bf k}))\Big] 
v_x^{\lambda,\lambda^\prime}({\bf k})v_x^{\lambda^\prime,\lambda}({\bf k})
\delta\big(E_\lambda({\bf k})-E_{\lambda^\prime}({\bf k})-\hbar\omega\big)\nonumber\\
& = & - \frac{e^2}{2\pi^2\omega}\int d^3  k \Big[f(E_{l}({\bf k}))-f(E_{h}({\bf k}))\Big]
v_x^{l, h}({\bf k})v_x^{ h,l}({\bf k})\delta(E_{l}({\bf k})-E_{ h}({\bf k})-\hbar\omega)
+ ( h\longleftrightarrow l).
\end{eqnarray}
\end{widetext}
It is to be noted that a factor 4 has been multiplied to obtain Eq. (\ref{OpT}). 
This is the results of the 2-fold degeneracy of the light and heavy holes. 
Using the following results 
$ v_x^{ h,l}({\bf k}) = [v_x^{l,h}]^* = - \sqrt{3}\gamma_2 (\hbar k/m_0)
\big[\cos\theta\cos\phi+i\sin\phi\big]$,
we can re-write Eq. (\ref{OpT}) as 
\begin{widetext}
\begin{eqnarray} \label{opcon-final}
\textrm{Re} \, [\sigma_{xx}(\omega)] & = & \frac{3e^2}{2\pi^2\omega}\gamma_2^2 
\frac{\hbar^2}{m_0^2}\int d^3k \,  
k^2(\cos^2\theta\cos^2\phi+\sin^2\phi)\delta\Big(\frac{2\gamma_2\hbar^2k^2}{m_0}-\hbar\omega\Big)
\Big[f(E_{ h}(k))-f(E_{l}(k))\Big]\nonumber\\
& = & \frac{e^2}{h} k_{\omega} \big[f(E_{ h}(k_\omega))-f(E_{l}(k_\omega))\big],
\end{eqnarray}
\end{widetext}
where $k_\omega=\sqrt{m_0\omega/2\gamma_2\hbar}$. 
An alternate derivation of the optical conductivity using the Green's function 
technique is given in Appendix.

At $T=0$ K, the above equation can be written as
\begin{eqnarray}
\textrm{Re} \, [\sigma_{xx}(\omega)] & = & \frac{e^2}{h} k_{\omega} \Theta(\hbar \omega - \epsilon_l)
\Theta(\epsilon_h - \hbar\omega),
\end{eqnarray}
where  $\epsilon_{h/l}=2\gamma_2 [\hbar k_{\rm f}^{h/l}]^2/m_0$ and $\Theta(x) $ 
is the unit step function.

In order to have interband transitions from heavy hole band to
light hole band at $T=0$, the photon energy must follow the 
inequality $ 2\gamma_2 [\hbar k_{\rm f}^l]^2/m_0 \leq \hbar \omega 
\leq 2\gamma_2 [\hbar k_{\rm f}^h]^2/m_0 $. 
With this, the magnitude of the optical conductivity at the left and right  edges
can be simply expressed as
\begin{eqnarray}
\textrm{Re} \, [ \sigma_{xx}  (\omega_{L/R}) ] = \frac{e^2}{h} k_{\rm f}^0
\frac{\sqrt{\gamma_1\pm2\gamma_2}}{[(\gamma_1+2\gamma_2)^{3/2} +
(\gamma_1-2\gamma_2)^{3/2}]^{1/3}} \nn,
\end{eqnarray}
where $ \omega_{L/R} = \epsilon_{h/l}/\hbar = 2\gamma_2 \hbar [k_{\rm f}^{h/l}]^2/m_0 $.
The optical band width (region in which $\sigma_{xx}(\omega)$
remains finite) is given by
\begin{eqnarray}
\Delta_{\rm op} = \frac{(4\gamma_2)^2E_{\rm f}^0}{[(\gamma_1+2\gamma_2)^{3/2} +
(\gamma_1-2\gamma_2)^{3/2}]^{2/3}}.
\end{eqnarray}
The band width $\Delta_{\rm op}$ and $ \textrm{Re} \, [ \sigma_{xx} (\omega_{L/R})] $ 
goes as $n_h^{2/3}$ and $ n_h^{1/3} $, respectively.

In Fig. 5, we plot the optical conductivity versus photon energy for three different
III-V semiconductors at $T=0$.

\begin{figure}[htbp]
\begin{center}\leavevmode
\includegraphics[width=92mm,height=70mm]{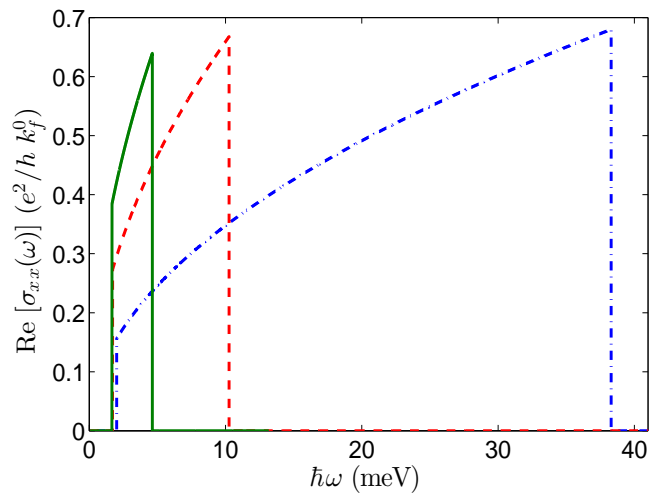}
\caption{Plots of the absorptive part of the optical conductivity 
(in units of $(e^2/h$ $k_f^0)$) versus photon energy for three 
different semiconductors: AlSb (solid green), GaAs (dashed red) 
and InAs (dashed-dot blue).}
\label{Fig1}
\end{center}
\end{figure}

\begin{table}[htbp]
\begin{center}
\centering
\caption{Values of $\tau_{h/l}, \epsilon_{h/l} $ and the 
Drude conductivity (in units of ($\Omega^{-1}$ $\mu$m$^{-1}$) 
along with the Luttinger parameters for six different III-V semiconductors.}
  \begin{tabular}{|c|c|c|c|c|c|c|c|}
  \addlinespace
  \toprule
\hline\hline
III-V & $\gamma_1$ & $\gamma_2 $ & $ \sigma$ &
$ \tau_{l} $ (ns) & $ \tau_h $ (ns)& $\epsilon_h $ (meV)& $\epsilon_l$
(meV)  \\\hline

GaAs & 7  & 2.5  & 15.703 & 0.062 & 0.571 &10.262 &1.710 \\
 InAs & 20  & 9   & 17.383 & 0.025 & 0.619 &38.295 & 2.015\\
 InSb & 35  & 15   & 36.457 & 0.057 & 0.513 &63.439 &4.879  \\
 AlSb & 5.24 & 1.23 & 16.221 & 0.115 &0.435  & 4.628 & 1.671 \\
 AlAs & 3.84 & 1.71 & 11.439 & 0.021 & 0.185 & 7.267 & 0.420\\
 GaSb & 13.4 & 4.7 & 26.759 & 0.091 & 0.472 & 19.228 & 3.373\\
  \hline
  \end{tabular}
\label{tab:addlabel}
\end{center}
\end{table}

\begin{figure}[htbp]
\begin{center}\leavevmode
\includegraphics[width=93mm,height=75mm]{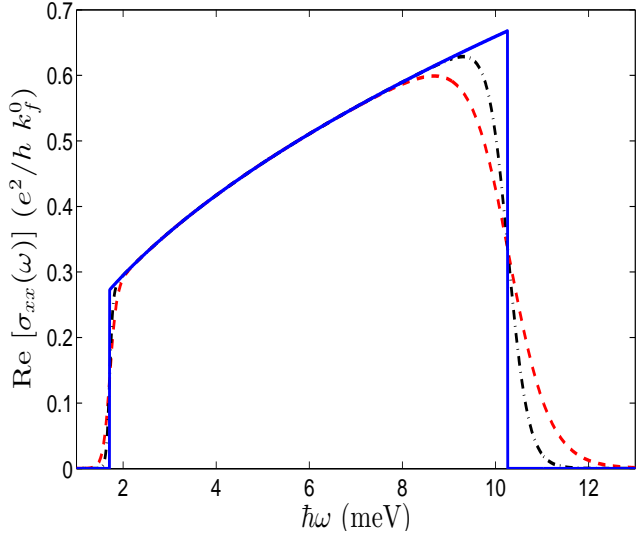}
\caption{Plots of the absorptive part of the optical conductivity versus
photon energy for GaAs at three different temperatures: $T=0$ K (solid blue),
$T = 0.5$ K (dot-dashed green) and  $T=1 $ K (dashed red).}
\label{Fig1}
\end{center}
\end{figure}

We show the optical conductivity at three different temperature in Fig. 6. 
It is easy to see that at the two edges the conductivity is 
${\rm Re} \, [\sigma_{xx} (\omega_{L/R})]/2$.
The location of the onset of optical transition and the magnitude of
the optical conductivity at this location do not change appreciably with the
temperature. The Luttinger parameters $\gamma_1$ and $ \gamma_2$ can be obtained from the
onset energy $\epsilon_l $ and  
$ {\rm Re} \, [\sigma_{xx}(\omega_L)] $. Thus, an approximate values of 
the Luttinger parameters can be obtained from the experimental measurement 
of the optical conductivity. It should be noted here that a precise determination 
of the Luttinger parameters requires a proper analysis of the 
$6 \times 6$ Luttinger Hamiltonian.

\section{Conclusion}
In this work, we have presented detailed analysis of the electrical and optical 
conductivities of hole gas in $p$-doped bulk III-V semiconductors described by the
$4 \times 4$ Luttinger Hamiltonian. The exact analytical expressions of the Drude conductivity,
inverse relaxation times for various impurity potentials, Drude weight and 
the optical conductivity are obtained.
We find that the back scattering is completely suppressed due to the 
helicity conservation of the heavy hole and light hole states.
The variation of the relaxation time with energy of the hole states is
different from the Brooks-Herring formula for electron gas in $n$-doped
semiconductors.
It is shown that the relaxation time of heavy holes is much larger than that 
of the light holes for long-range impurity potentials and vice-versa for 
the short-range impurity potential.
Note that our results are valid as long as the Fermi energy is smaller 
than the split-off energy. For more accurate results, one need to consider 
the $6 \times 6 $ Luttinger
Hamiltonian. The effective masses of the HH and LH 
bands described by the $6 \times 6$ Luttinger Hamiltonian will be different from
the HH and LH bands described by the $4 \times 4$ Hamiltonian. 
The effective band mass appears in the density of states
and consequently in the inverse relaxation time as well as in the Drude conductivity.
There will be a quantitative change in the inverse relaxation time 
and the Drude conductivity if we consider $6 \times 6 $ Luttinger Hamiltonian.

The Drude weight has a linear density dependency even with the 
non-zero spin-orbit coupling, due to $k^2$ dependence of the spin-orbit coupling. 
The finite-frequency optical conductivity is having $\sqrt{\omega}$ dependence. 
The onset energy for triggering optical transition and the amplitude of the optical
conductivity at the onset energy depend on the Luttinger parameters $\gamma_1 $ and 
$\gamma_2$. Therefore, the Luttinger parameters $\gamma_1 $ and $\gamma_2$  can be 
determined approximately from the optical measurements.
The values of the Luttinger parameters, $\tau_{h/l}$ at the Fermi energy for the
Coulomb-type impurity potential, Drude conductivity at $T=0$ for the Coulomb-type
impurity potential, onset energy $\epsilon_h$ and 
offset energy $\epsilon_l$  for the optical transition for six
different III-V semiconductors are tabulated in Table 1.

\appendix

\section{Alternative derivation of the optical conductivity}
Here we shall provide alternative derivation of the optical 
conductivity.
Using the Kubo formula, the optical conductivity 
can also be written as 
\begin{eqnarray}\label{kubo}
& &\sigma_{xx}(\omega)=-\frac{e^2}{i\omega}\frac{1}{(2\pi)^3}
\int_0^\infty\int_0^\pi\int_0^{2\pi}k^2 \sin\theta  dk d\theta d\phi \nonumber\\
&\times& T \sum_s \textrm{Tr} \langle \hat{v}_x \hat{G}({\bf k},\omega_s) 
\hat{v}_x \hat{G}({\bf k},\omega_s+\omega_n) \rangle_{i\omega_n 
\rightarrow \omega + i\delta}.
\end{eqnarray}
Here $T$ is the temperature, $s$ and $n$ are integers, $\omega_n=(2n+1) \pi T$ 
and $\omega_s = 2s \pi T$ are the fermionic and bosonic Matsubara frequencies, respectively.

The Green's function of the Luttinger Hamiltonian [Eq. (\ref{hamil})] is given by
\begin{widetext}
\begin{eqnarray}\label{green}
\hat{G}({\bf k},\omega_n)&=&\sum_\lambda\begin{bmatrix}
-\frac{1}{2}-\lambda\frac{1+3\cos2\theta}{8}& -\lambda 
\frac{\sqrt{3}}{4}\sin2\theta \e^{-i\phi} & -\lambda 
\frac{\sqrt{3}}{4}\sin^2\theta \e^{-i2\phi} &0\\
 -\lambda\frac{\sqrt{3}}{4}\sin2\theta \e^{i\phi} & 
-\frac{1}{2}+\lambda\frac{1+3\cos2\theta}{8} & 0 &  
- \lambda\frac{\sqrt{3}}{4}\sin^2\theta \e^{-i2\phi} \\
-\lambda\frac{\sqrt{3}}{4}\sin^2\theta \e^{i2\phi} & 0& 
-\frac{1}{2}+\lambda\frac{1+3\cos2\theta}{8} &  
-\lambda\frac{\sqrt{3}}{4}\sin2\theta \e^{-i\phi}\\
  0 &  -\lambda\frac{\sqrt{3}}{4}\sin^2\theta \e^{i2\phi} &  
-\lambda\frac{\sqrt{3}}{4}\sin2\theta \e^{i\phi} & -\frac{1}{2} 
-\lambda\frac{1+3\cos2\theta}{8}
\end{bmatrix}G_0^\lambda({\bf k},\omega_n),
\end{eqnarray}

\end{widetext}
where $G_0^\lambda({\bf k},\omega_n)=1/(E_{\lambda} - \mu - i\hbar\omega_n).$
Using Eqs. (\ref{vx}) and (\ref{green}), one can obtain
\begin{widetext}
\begin{eqnarray}
& &\textrm{Tr}\langle\hat{v}_x\hat{G}({\bf k},\omega_s)\hat{v}_x 
\hat{G}({\bf k}, \omega_s+\omega_n)\rangle = 
\Big\{ [2 \gamma_1^2(1+\lambda\lambda^\prime) 
- 8\gamma_1\gamma_2 (\lambda + \lambda^\prime)] \cos^2\phi 
+ \gamma_2^2\big[13 +
(7 \lambda\lambda^\prime +1)\cos2\phi -5 \lambda \lambda^{\prime}\big] \nonumber\\
& + & 2 \big[4\gamma_1\gamma_2(\lambda+\lambda^\prime) - 
\gamma_1^2(1+\lambda\lambda^\prime) - \gamma_2^2(1+ 7\lambda\lambda^\prime)\big] 
\cos2\theta \cos^2\phi 
\Big\}  \Big(\frac{\hbar k}{2 m_0}\Big)^2
G_0^\lambda({\bf k},\omega_n)
G_0^{\lambda^{\prime}}({\bf k},\omega_n+\omega_s).
\end{eqnarray}
\end{widetext}
The identity
\begin{eqnarray}
&&T \sum_s \bigg[\frac{1}{i\hbar\omega_s+\mu-E_\lambda}
\frac{1}{i\hbar(\omega_n+\omega_s)+\mu-E_{\lambda^\prime} }\bigg]\nonumber\\
&=& 
\begin{cases}
\frac{f(E_\lambda)-f(E_{\lambda^\prime})}{i\hbar\omega_n-E_{\lambda^\prime}+E_{\lambda}},& \text{if } 
\lambda\neq \lambda^\prime\\
    0,              & \text{otherwise.}
\end{cases}
\end{eqnarray}
shows that there is no intra-band contribution to the optical conductivity.
Thus by keeping only the terms involving inter-band transitions, we have
\begin{eqnarray}\label{trace}
&& T \sum_s \textrm{Tr}\langle v_x \hat{G}({\bf k},\omega_s)v_x 
\hat{G}({\bf k},\omega_n+\omega_s) \rangle \nonumber\\
&=&6\Big(\frac{\gamma_2\hbar k}{m_0}\Big)^2
\Big[\cos^2\theta\cos^2\phi+\sin^2\phi\Big]\nonumber\\
&\times&\Big[\frac{f(E_h)-f(E_l)}{i\hbar\omega_n-E_l+E_h}\Big]+(E_h\leftrightarrow E_l).
\end{eqnarray}
It is to be noted that the second term turns out to be zero as 
a result of the conservation of energy.

Using the result of Eq. (\ref{trace}) in Eq. (\ref{kubo}), we have 
\begin{eqnarray}
& &\sigma_{xx}(\omega) = -\frac{e^2}{i\omega}\frac{1}{(2\pi)^3}
\int_0^\infty\int_0^\pi\int_0^{2\pi}k^2 dk \sin\theta d\theta d\phi\nonumber\\
&\times&6\Big( \frac{\gamma_2 \hbar k}{m_0}\Big)^2
\Big[\cos^2\theta\cos^2\phi+\sin^2\phi\Big]\nonumber\\
&\times&\Big[\frac{f(E_h)-f(E_l)}{i\hbar\omega_n-E_l+E_h}\Big]+(E_h\leftrightarrow E_l)_{i\omega_n
\rightarrow \omega+i\delta}.
\end{eqnarray}
On further simplification the absorptive part of the optical conductivity  reduces to
\begin{widetext}
\begin{eqnarray}
\textrm{Re} \, [ \sigma_{xx}(\omega) ] & = &\frac{3e^2}{2\pi^2\omega}\gamma_2^2
\frac{\hbar^2}{m_0^2}\int d^3k \, k^2(\cos^2\theta\cos^2\phi+\sin^2\phi)
\delta\Big(\frac{2\gamma_2\hbar^2k^2}{m_0}-\hbar\omega\Big)
\Big[f(E_{h}(k))-f(E_{l}(k))\Big]\nonumber\\
& = &\frac{e^2}{h}k_\omega\big[f(E_{h}(k_\omega))-f(E_{l}(k_\omega))\big].
\end{eqnarray}
\end{widetext}

\end{document}